\documentclass[12pt]{article}

\usepackage{authblk}
\usepackage{multibib}

\usepackage[superscript]{cite}
\usepackage[mathlines]{lineno}

\usepackage{times}

\usepackage[T1]{fontenc}
\usepackage{caption}
\usepackage{sectsty}
\sectionfont{\fontsize{13}{15}\selectfont}
\subsectionfont{\fontsize{12}{15}\selectfont}
\usepackage{nameref}
\usepackage{titlesec}

\usepackage[symbol]{footmisc}

\usepackage{lineno}
\usepackage{color}
\usepackage{bm}

\usepackage[english]{babel}
\usepackage{graphicx}
\usepackage{dcolumn}
\usepackage[utf8]{inputenc} 
\usepackage{bm}
\usepackage{hyperref}
\usepackage{mathrsfs}
\usepackage{longtable}
\usepackage{multirow,booktabs,dcolumn,tabularx}
\usepackage{float}

\usepackage{braket}
\usepackage{microtype} 
\usepackage[dvipsnames]{xcolor}

\usepackage{amsmath,amsfonts,amsthm,amssymb} 

\usepackage{dsfont}
\usepackage{comment}
\usepackage{physics}
\usepackage{booktabs}
\usepackage{pifont}
\usepackage{makecell}

\newcommand{\rv}{{\bf r}}
\newcommand{\iv}{{\bf i}}
\newcommand{\jv}{{\bf j}}
\newcommand{\up}{\uparrow}
\newcommand{\down}{\downarrow}
\newcommand{\la}{\langle} 
\newcommand{\ra}{\rangle} 

\usepackage{abstract}

\makeatletter
\renewcommand{\maketitle}{\bgroup\setlength{\parindent}{0pt}
\begin{flushleft}
  \textbf{\@title}\\

  \@author
\end{flushleft}\egroup
}
\makeatother

\begin{document} 

\title{\noindent {\bf \large Gapless superconductivity from extremely dilute magnetic disorder in 2H-NbSe$_{2-x}$S$_x$}}

\author[1]{\normalsize Jose Antonio Moreno}
\author[2,3]{\normalsize Merc\`e Roig} 
\author[1]{\normalsize V\'ictor Barrena}
\author[1]{\normalsize Edwin Herrera}
\author[4]{\normalsize Alberto M. Ruiz}
\author[4]{\normalsize Samuel Ma\~{n}as$-$Valero}
\author[1]{\normalsize Ant\'on Fente}
\author[1,5]{\normalsize Anita Smeets}
\author[6]{\normalsize Jazm\'in Arag\'on}
\author[6]{\normalsize Yanina Fasano}
\author[1]{\normalsize Beilun Wu}
\author[7]{\normalsize Maria N. Gastiasoro}
\author[4]{\normalsize Eugenio Coronado}
\author[4]{\normalsize Jos\'e J. Baldov\'i}
\author[2]{\normalsize Brian M. Andersen}
\author[1]{\normalsize Isabel Guillam\'on}
\author[1]{\normalsize Hermann Suderow}

\affil[1]{\small \it Laboratorio de Bajas Temperaturas y Altos Campos Magn\'eticos, Departamento de F\'isica de la Materia Condensada, Instituto Nicol\'as Cabrera and Condensed Matter Physics Center (IFIMAC), Unidad Asociada UAM-CSIC, Universidad Aut\'onoma de Madrid, E-28049 Madrid, Spain}
\affil[2]{\small \it Niels Bohr Institute, University of Copenhagen, DK-2100 Copenhagen, Denmark}
\affil[3]{\small \it Department of Physics, University of Wisconsin–Milwaukee, Milwaukee, Wisconsin 53201, USA} 
\affil[4]{\small \it Instituto de Ciencia Molecular (ICMol), Universidad de Valencia, Catedr\'atico Jos\'e Beltr\'an 2, 49680, Paterna, Spain} 
\affil[5]{\small \it MESA+ Institute for Nanotechnology, University of Twente, 7500 AE, Enschede, The Netherlands}
\affil[6]{\small \it Low Temperatures Lab, Centro Atómico Bariloche, CNEA, 8400 Bariloche, Argentina}
\affil[7]{\small \it Donostia International Physics Center, 20018 Donostia-San Sebastian, Spain}

            \date{}

\baselineskip24pt

\maketitle 


\noindent{\bf Most superconducting materials exhibit a vanishing density of states at the Fermi level and Anderson’s theorem posits that the superconducting gap is robust against nonmagnetic disorder. Although dilute magnetic impurities lead to localized in-gap states, these states typically have no bearing on the material's bulk superconducting properties. However, numerous experiments reveal a finite density of states at the Fermi level in systems with an apparently negligible number of magnetic impurities. Here, using scanning tunneling microscopy and self-consistent Bogoliubov-de Gennes calculations, we find that gapless superconductivity emerges in 2H-NbSe$_{2-x}$S$_x$ at remarkably low magnetic impurity concentrations. Furthermore, our density functional theory calculations and in-gap quasiparticle interference measurements demonstrate that the Se-S substitution significantly modifies the band structure. This modification favours nesting and dictates the in-gap scattering for $x>0$, in stark contrast to the dominant charge density wave interactions in pure 2H-NbSe$_2$. Our findings reveal an unusual superconducting response to disorder and highlight the importance of incorporating material-specific band structures in the understanding of a superconductor's response to even very low concentrations of magnetic impurities.}

\section*{Introduction}
Quantum materials are characterized by coupled quantum degrees of freedom which can give rise to a plethora of emergent ordered states, including, for example, superconductivity and charge density waves. Disorder can act as an important probe for such many-body phases through the detailed response to local spatially modulated perturbations. In conventional $s$-wave superconductors, nonmagnetic impurities generally do not significantly alter superconductivity due to the preservation of time-reversed Kramer's pairs\,\cite{Anderson1959}. This explains the prevalence of superconductivity in numerous disordered metals\,\cite{Anderson1959,PhysRevLett.108.017002,PhysRevLett.98.027001,PhysRevB.90.134513,PhysRevB.72.060502,PhysRevB.98.184510}. Despite this, a persistent observation in many superconductors -- even those with seemingly negligible amounts of magnetic impurities -- is a nonzero density of states (DOS) at the Fermi level\,\cite{Rubio-Verdu2020,Nayak2021,Zhao2019,Wan2023,Naritsuka2025,Postolova2020,PhysRevResearch.4.023241,PhysRevB.87.094502,PhysRevLett.96.027003}. While gapless superconductivity is often associated with unconventional superconducting behaviors, like magnetic order\,\cite{Deng2024,Deng20242,PhysRevLett.96.027003}, finite-momentum superconductivity\,\cite{Zhu2021,Yuan2018}, and with Bogoliubov Fermi surfaces\,\cite{Zhu2021,ohashi2024,Lee2023,10.21468/SciPostPhysCore.5.1.009} or Majorana bound states\,\cite{Papaj2021}, its connection to the normal phase electronic bandstructure and the electronic correlations of the host material is rarely explored. Although this relationship is fundamental in highly disordered and amorphous superconductors\cite{PhysRevLett.98.027001,PhysRevB.90.134513,PhysRevB.72.060502,PhysRevB.98.184510,Postolova2020,RevModPhys.80.1355,PhysRevLett.108.017002,Zhao2019}, its relevance in single crystals with substitutional disorder remains unclear.

Here we use atomic-scale scanning tunneling microscopy (STM) to investigate 2H-NbSe$_{2-x}$S$_{x}$ ($0\leq x\leq0.8$) doped with very low concentrations of randomly distributed Fe impurities ($n\approx$ 0.02 at.\,\%). We find that minute amounts of Fe disorder in conjunction with the S substitutional disorder  cooperate to produce a gapless superconducting state. By comparing experimental results with material-specific theoretical simulations of in-gap magnetic bound states and nonmagnetic disorder, we demonstrate that a comprehensive understanding of the resulting finite in-gap DOS necessarily involves incorporating the changes in the normal state band structure induced by the nonmagnetic S substitution. In Fig.\,\ref{fig:Schematics} we provide a schematic illustration of the emergence of localized impurity states and gapless superconductivity from an interplay between such magnetic in-gap bound states and substitutional disorder.

\section*{Results}

Figures\,\ref{fig:YSRtoAG}{\bf a-c} display tunneling conductance maps from two distinct regions of a 2H-NbSe$_{1.8}$S$_{0.2}$ sample (we provide experimental details in Methods, Sections A and B and Extended Data Table \ref{tab:caracterizacion}). The random distribution of Fe impurities, marked by black dots in Fig.\,\ref{fig:YSRtoAG}{\bf a,b}, leads to substantial local variations of the impurity density. Despite the large inter-impurity distances, well above the superconducting coherence length of $\xi=7$ nm in 2H-NbSe$_{1.8}$S$_{0.2}$, the tunneling conductance far from impurities consistently exhibits finite conductance at zero bias, deviating from the typical zero in-gap conductance of clean BCS superconductors. It is remarkable that such extremely dilute concentrations of introduced magnetic moments, corresponding to 1 moment per $\sim 3000$ unit cells, are capable of producing gapless superconductivity throughout the field of view. 

To understand these experimental findings, we present in Fig.\,\ref{fig:YSRtoAG}{\bf d,e} the calculated local density of states (LDOS) maps obtained from selfconsistent Bogoliubov-de Gennes simulations. These results are based on a microscopic model that includes randomly placed point-like dilute magnetic moments on a triangular lattice with tight-binding hopping parameters that reproduce the salient features of the band structure of 2H-NbSe$_{2-x}$S$_{x}$ (all details are provided in Methods, Section C and Fig.~\ref{fig:S1} and \ref{fig:DOSnorm}). The obtained average LDOS shown in Fig.\,\ref{fig:YSRtoAG}{\bf f}, which is representative of the typical LDOS curves in-between magnetic impurities, successfully reproduces the experimental observation of an enhanced LDOS at the Fermi level. We return to a detailed discussion of the quantitative aspects of the comparison to experiments in the Discussion section.

Our findings are consistent across multiple fields of views and for various 2H-NbSe$_{2-x}$S$_{x}$ compositions ($0.2\leq x \leq 0.8$). The experimental tunneling conductance far from magnetic impurities is presented in Fig. \ref{fig:gaplessdependence}{\bf a} for 2H-NbSe$_{1.8}$S$_{0.2}$ as a function of magnetic impurity concentration $n$; and in Fig. \ref{fig:gaplessdependence}{\bf c} for 2H-NbSe$_{2-x}$S$_{x}$ as a function of S content ($x$). The corresponding calculated LDOS are shown in Figs.\,\ref{fig:gaplessdependence}{\bf b} and {\bf d}. As seen, there is qualitative agreement with the experimental results, both in terms of magnetic and nonmagnetic disorder effects. Firstly, as expected, the enhanced concentrations of magnetic moments pushes spectral weight into the superconducting gap, as seen from Figs.\ref{fig:gaplessdependence}{\bf a} and {\bf b}. Secondly, a key result is that increased substitutional (nonmagnetic) disorder clearly reduces the gap and amplifies the influence of magnetic impurities on the low-energy LDOS, as seen from Figs.\,\ref{fig:gaplessdependence}{\bf c} and {\bf d}. For additional results, see Figs.~\ref{fig:DOSnorm},\ref{fig:S2},\ref{fig:FOV},\ref{fig:azufres} and Supplementary Information Sections A-C, E. An understanding of the latter effect requires detailed knowledge of the general trends in the evolution of the normal state band structure with S substitution, an important topic to which we turn next.   

We now show that the scattering patterns produced inside the gap by magnetic impurities in 2H-NbSe$_2$ and 2H-NbSe$_{1.8}$S$_{0.2}$ present compelling evidence for significant changes induced by S substitution in the electronic band structure. Tunneling conductance maps reveal spatial modulations of the LDOS, a characteristic feature of localized magnetic impurities. The wavelength of these modulations is dictated by the dominant features in the band structure responsible for pair-breaking interaction\,\cite{Yazdani1997,Menard2015,victorvortex,Liebhaber2020,PhysRevLett.120.167001}. Figures\,\ref{fig:qpifft}{\bf b-d} show the Fourier transforms of these maps at selected in-gap bias voltages for 2H-NbSe$_2$. In contrast, Figs.\,\ref{fig:qpifft}{\bf f-h} display the corresponding data for 2H-NbSe$_{1.8}$S$_{0.2}$ (corresponding topographies are shown in Fig. \ref{fig:qpifft}{\bf a,e}. Full data and real space maps are provided in Fig.\,\ref{fig:qpiall}, see also Supplementary Information Section C). In 2H-NbSe$_2$, we observe LDOS modulations at reciprocal wave vectors corresponding to the atomic $q_{at}$ and charge density wave (CDW) $q_{CDW}$ periodicities, consistent with previous studies \cite{Menard2015,PhysRevB.100.014502,Kezilebieke2018,Liebhaber2020,PhysRevB.100.014502,victorvortex}. However, 2H-NbSe$_{1.8}$S$_{0.2}$ exhibits additional, distinct features. At bias voltages of $0.7\,$mV (Fig.\,\ref{fig:qpifft}{\bf g}), two new wave vectors, $q_1$ and $q_2$, emerge at the Brillouin zone boundary, clearly separated from the CDW wave vectors. Increasing the bias to $1\,$mV (Fig.\,\ref{fig:qpifft}{\bf h}) leads to a decrease in the intensity of these modulations, accompanied by the emergence of two additional wave vectors: $q_3$, rotated by 30$^{\circ}$ relative to the atomic and CDW lattices, and $q_4$, forming a near-circular scattering pattern.

To understand the origin of the new in-gap spatial oscillations, we performed density functional theory (DFT) calculations on two equivalent five-unit-cell slabs along the c-axis for both 2H-NbSe$_2$ and 2H-NbSe$_{1.8}$S$_{0.2}$ (see Methods Section D for details). The results, along with the three-dimensional Brillouin zone, are presented in Figs\,.\,\ref{fig:qpifft}{\bf i,j,l,m} and Fig.\,\ref{fig:qpifft}{\bf k}, respectively. In pure 2H-NbSe$_2$ (Fig.\,\ref{fig:qpifft}{\bf i}), the Fermi surface features a nearly spherical three-dimensional Se-derived central pocket at the Brillouin zone center. The Nb-derived tubular-like pockets show substantial $k_z$ warping. This difference in band dispersion between the Brillouin zone center ($\Gamma-K-M$) and at the upper basal planes ($A-H-L$) is clear in Fig.\,\ref{fig:qpifft}{\bf l}, where the corresponding energy dispersion curves do not overlap.

Upon S substitution in 2H-NbSe$_{1.8}$S$_{0.2}$ (Fig.~\ref{fig:qpifft}{\bf j}), we observe a slight reduction in the size of the central Se/S-derived pocket. More significantly, the Nb-derived sub-bands now exhibit well-separated Fermi surface sheets within the three-dimensional Brillouin zone. The crucial change occurs because S substitution substantially diminishes the $k_z$ warping. The varying S concentration for each plane within the slab effectively decouples the electronic band structure between layers, leading to sub-bands with a pronounced two-dimensional character. Consequently, the band dispersion in 2H-NbSe$_{1.8}$S$_{0.2}$ (Fig.\ref{fig:qpifft}{\bf m}) shows nearly identical dispersion at both the Brillouin zone center ($\Gamma-K-M$) and the upper basal planes ($A-H-L$) within each sub-band.

We can correlate the wave vectors observed in the in-gap scattering patterns (Fig.\,\ref{fig:qpifft}{\bf g,h}) with the features of the calculated band structures (Figs.\,\ref{fig:qpifft}{\bf l,m}). Specifically, the wave vector $q_3$ connects opposite, nearly flat portions of the hexagonal Nb-derived Fermi surface pocket centered at $\Gamma$. This is consistent with the main nesting wave vector of the band structure\,\cite{Johannes2006}. Furthermore, the wave vector $q_4$ (green in Fig.\,\ref{fig:qpifft}{\bf j}) is related to the Se/S-derived central pocket. Finally, the wave vectors $q_1$ and $q_2$ bridge the M and L points of the Brillouin zone, which are regions where the band structure exhibits significant energy curvature.

Consequently, S substitution in 2H-NbSe$_{2-x}$S$_{x}$ profoundly modifies the normal state electronic band structure compared to pure 2H-NbSe$_2$. The electronic features associated with the CDW become less influential in their interaction with magnetic impurities. Instead, other electronic states distributed throughout the entire band structure now dominate the scattering. Concurrently, we observe a decrease in the superconducting critical temperature $T_c$ (Fig.\,\ref{fig:TC}) and the suppression of the CDW with increasing sulfur concentration $x$ (Fig.\,\ref{fig:Topos} and Supplementary Information Section D).

\section*{Discussion}

Although in-gap Yu-Shiba-Rusinov (YSR)~\cite{Yu1965,Shiba1968,Rusinov1969} states induced by magnetic impurities have been widely studied in pure 2H-NbSe$_{2}$, these investigations consistently report a superconducting gap that remains fully open away from magnetic impurities~\cite{Menard2015,PhysRevB.100.014502,Kezilebieke2018,Liebhaber2020,PhysRevB.100.014502}. In contrast, gapless superconductivity in conventional systems typically emerges only at concentrations of magnetic impurities significantly higher than found here, on the order of several atomic percent as opposed to a fraction of a per mille~\cite{PhysRevB.8.1038,PhysRevB.78.174518,Iavarone_2011,Iavarone_2010}. We stress that the gapless state found here appears despite the fact that for all $x>0$, the YSR bound states centered on magnetic impurities remain rather isolated and exhibit characteristics similar to those in 2H-NbSe$_2$, including a pronounced electron-hole asymmetry and an exponential decrease with distance (see Figs.\,\ref{fig:S2},\ref{fig:FOV},\ref{fig:azufres},\ref{fig:asymmetrydecay} and Supplementary Information section A-C, E). 

There is a quantitative discrepancy between the experimental data and the theoretical findings: our microscopic calculations require higher magnetic impurity concentrations to reproduce the experimental effects. For example, the experimental data in Fig.~\ref{fig:YSRtoAG}{\bf b},{\bf c} (0.032\% Fe concentration) yields results comparable to calculations in Fig.~\ref{fig:YSRtoAG}{\bf e},{\bf f}, which correspond to a 0.229\% magnetic impurity concentration. Despite this discrepancy, however, even this value is remarkably low for inducing gapless superconductivity~\cite{Shiba1968}. The origin of the extreme sensitivity to magnetic impurities lies in the spatial extension of the YSR bound state wavefunctions. Their extent is often longer for reduced dimensionality, but also the normal state band structure can play an important role~\cite{Balatsky2006,Uldemolins2022,Uldemolins2024}. In the present case, for the clean system, the YSR states exhibit very long tails (see Figs.\,\ref{fig:S1}-\ref{fig:S2}), which can facilitate the finite LDOS far from impurities. Remarkably, even though the long YSR tails are smeared out by the nonmagnetic substitutional disorder (see Fig. \ref{fig:S2}), the two types of disorder cooperate in promoting the gapless state as seen in Fig.~\ref{fig:YSRtoAG} and Fig.\,\ref{fig:DOSnorm}. The quantitative discrepancy between theory and experiment suggests that our model may underestimate the cooperative influence of Se-S substitutional disorder, or that it indicates the relevance of additional degrees of freedom, such as enhanced RKKY couplings from electronic correlations~\cite{Moroni2017}, CDW feedback effects, variations in the electron-phonon coupling, or additional band structure modifications by the S doping unaccounted for in the lattice model simulations.

Modifications of the normal state band structure are particularly important for understanding the $x$ dependence of the LDOS in 2H-NbSe$_{2-x}$S$_{x}$ in Fig.~\ref{fig:gaplessdependence}{\bf c,d}. In the theoretical model, the normal state DOS at the Fermi level is reduced with (nonmagnetic) $x$ disorder. This has two important effects; first, it reduces the overall superconducting gap scale, in agreement with the narrowing of the coherence peaks seen in Fig.~\ref{fig:gaplessdependence}{\bf c,d}. Second, the reduced DOS naturally enhances the normalized zero energy tunneling conductance, also in agreement with the results displayed in Fig.~\ref{fig:gaplessdependence}{\bf c,d}. This latter effect is detailed in Fig.~\ref{fig:DOSnorm}. Thus, in conclusion, the observed non-zero LDOS at the Fermi level stems the influence of nonmagnetic disorder in both the LDOS of magnetic YSR states, and the normal state LDOS.

We have also shed light on the complex relationship between superconductivity and CDW order in 2H-NbSe$_2$  (see Fig.\,\ref{fig:Topos} and Supplementary Information Section D for the evolution of the CDW with $x$). While a decrease in $T_c$ concurrent with CDW suppression has been observed in electron-irradiated 2H-NbSe$_2$ \cite{Cho2018}, other arguments suggest the CDW should have a limited impact on superconducting properties\,\cite{Das2023,Tissen2013,PhysRevResearch.2.043392}. In this context, we propose that disorder plays a crucial role in modifying electronic interactions. It likely shifts the dominant interaction from a specific wave vector (the CDW wave vector), where phonons are significant\,\cite{Johannes2006}, towards electron-electron interactions. The latter involve band structure features, such as the nesting feature at $q_3$,  which occur along different reciprocal space directions and do not participate in the formation of the CDW\,\cite{Johannes2006}.

\section*{Conclusions}

In summary, our study reveals a continuous evolution towards gapless superconductivity in an $s$-wave superconducting system, 2H-NbSe$_{2-x}$S$_x$. While material-specific modeling that includes both magnetic and nonmagnetic disorder finds a very low concentration of magnetic moments required for inducing a gapless state, additional effects are necessary to quantitatively capture the observations. We have exposed how modifications in the normal state band structure play an important role in facilitating an early onset of gapless superconductivity. The emergence of this state at remarkably low magnetic impurity concentrations highlights a cooperative interplay between Yu-Shiba-Rusinov bound states and nonmagnetic substitutional disorder on the resulting magnetic pair breaking effect in this superconducting material. 

\clearpage

\section*{Acknowledgements}

This work was supported by the Spanish Research State Agency (PID2023-150148OB-I00, PID2020-114071RB-I00, PDC2021-121086-I00, TED2021-130546B\-I00, CEX2023-001316-M and CEX2024-001467-M), Comunidad de Madrid through project TEC-2024/TEC-380 “Mag4TIC”, the EU through grant agreement No 871106, by the European Research Council PNICTEYES grant agreement 679080 and ERC-2021-StG 101042680 2D-SMARTiES. J.A.M, E.H., I.G., and H.S. acknowledge the “QUASURF” project [SI4/PJI/2024-00199] funded by the Communidad de Madrid through the direct grant agreement for the promotion and development of research and technology transfer at the Universidad Autónoma de Madrid. J.J.B acknowledges the Generalitat Valenciana (grant CIDEXG/2023/1) and A.M.R. thanks the Spanish MIU (Grant No FPU21/04195). The calculations were performed on the Tirant III cluster of the Servei d’Inform\'atica of the University of Val\`encia. We acknowledge collaborations through EU program Cost CA21144 (Superqumap.eu). We acknowledge SEGAINVEX for design and construction of cryogenic equipment and SIDI for support in sample characterization. M.N.G is supported by the Ramon y Cajal Grant RYC2021-031639-I funded by MCIN/AEI/ 10.13039/501100011033. M.R. acknow\-ledges support from the Simons Foundation Grant SFI-MPS-NFS-00006741-02.

\section*{Author contributions.} \
\noindent JAM and VB performed the experiments, with the supervision of EH, IG and HS. MR did the calculations with the supervision of MNG and BMA. Samples were synthesized and characterized by SMV and EG. DFT calculations were performed by AMR supervised by JJB. Vortex lattice experiments in doped samples were made by AF, JAM and JA, with the input of YF. VB and AS worked on the image treatment and interpretation. The study was devised by IG, HS, MNG and BMA. The paper was written by JAM, MR, AMR, JJB, MNG, BMA, IG and HS with input from all authors.

\section*{Competing Interests} \
\noindent The authors declare no competing interests.

\section*{Data availability.} \
\noindent The data generated in this study have been deposited in the osf.io database under accession code ....

\clearpage

\begin{figure*}[ht]
	\centering
		\includegraphics[width=\linewidth]{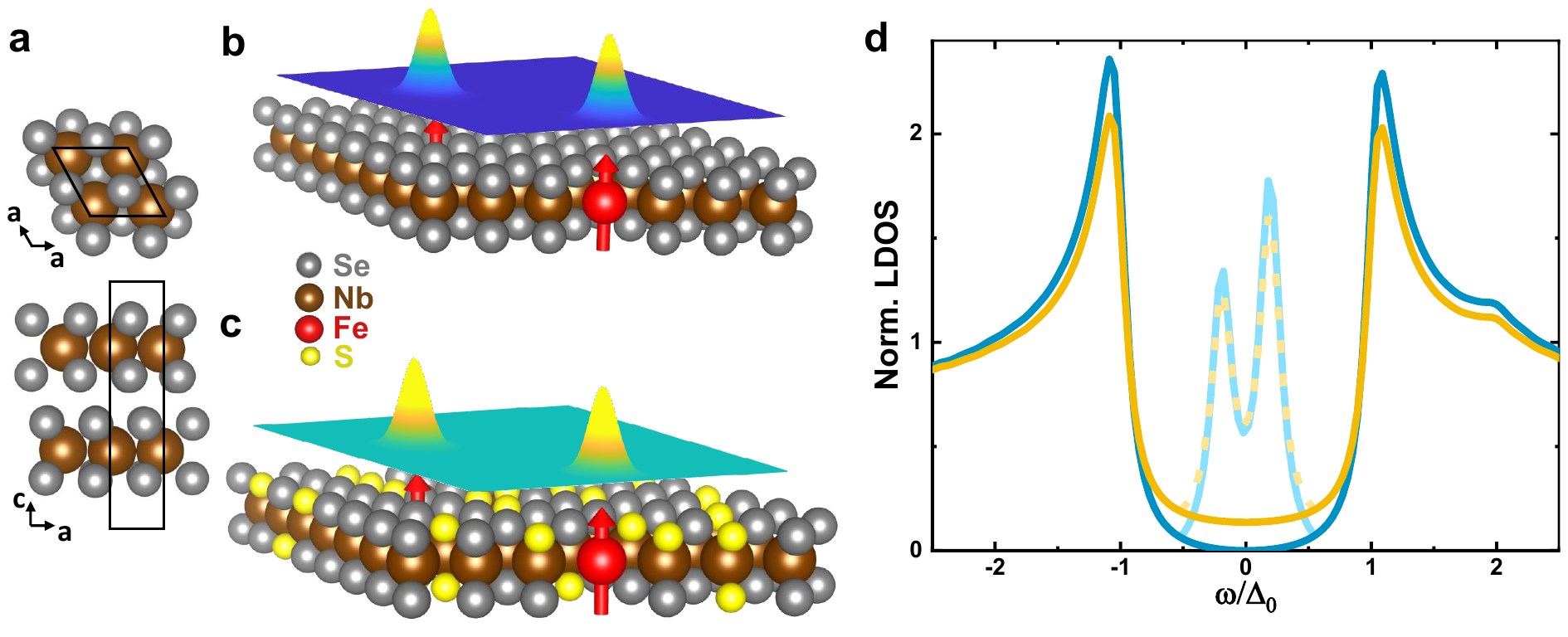} 
	\caption{\ \textbf{Emergence of gapless superconductivity in the 2H-NbSe$_{2-x}$S$_x$ system with dilute magnetic impurities}. {\bf a} Schematic of the 2H-NbSe$_2$ crystal structure: view of the aa-plane (ac-plane) in the top panel (bottom panel). The unit cell is indicated by the black line. Nb and Se atoms are brown and gray spheres, respectively. {\bf b} Schematic showing a single Se-Nb-Se layer of 2H-NbSe$_2$ containing two Fe magnetic impurities (red spheres with arrows). The color scale represents the local density of states (LDOS) at zero energy, from blue (zero LDOS) to yellow (LDOS peak at the impurity). {\bf c} Similar to {\bf b}, but for a 2H-NbSe$_{2-x}$S$_x$ layer ($x>0$) with S substitutional disorder (yellow spheres). {\bf d} Schematic representation of the LDOS as a function of $\omega$ normalized to the superconducting gap $\Delta_0$. Blue lines indicate pure 2H-NbSe$_2$: dark (light) blue for LDOS far from (at) a magnetic impurity. Ocre lines represent 2H-NbSe$_{2-x}$S$_x$ ($x>0$): dark (light) ocre for LDOS far from (at) an magnetic impurity. This panel highlights the modifications to the LDOS due to both Fe magnetic impurities and S substitution.}
		\label{fig:Schematics}
\end{figure*}

\begin{figure*}[ht]
	\centering
		\includegraphics[width=0.9\linewidth]{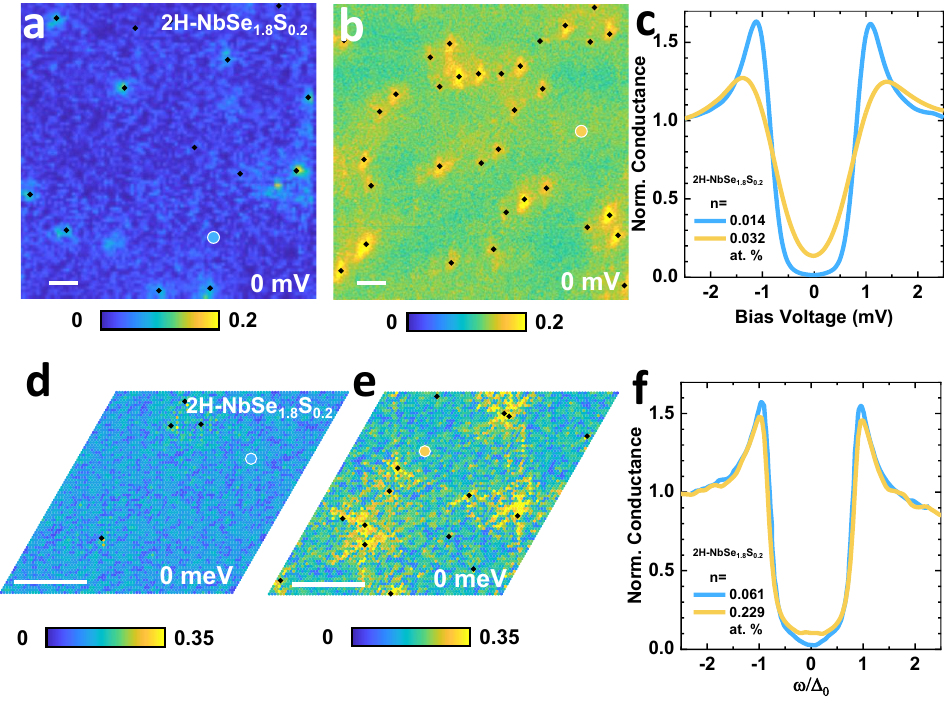} 
	\caption{\ \textbf{Gapless superconductivity in 2H-NbSe$_{1.8}$S$_{0.2}$.} {\bf a,b} Zero bias tunneling conductance maps in 2H-NbSe$_{1.8}$S$_{0.2}$ for low ({\bf a}, $n=0.014$ at.\,$\%$) and high Fe concentrations ({\bf b}, $n=0.032$ at.\,$\%$) obtained in different fields of view by STM. The temperature is $T/T_c\ll 0.05$. The white scale bars represent 10 nm in all maps. The color scale, ranging from zero (blue) to 0.2 (yellow) represents tunneling conductance normalized at bias voltages above the superconducting gap. Black dots mark the position of the Fe impurities. {\bf c} Normalized tunneling conductance curves taken at locations far from impurities (at the locations marked with circles in {\bf a,b}). {\bf d,e} Calculated LDOS for low ({\bf d}, $n=0.061$ at. $\%$) and high Fe concentrations ({\bf e}, $n=0.229$ at. $\%$). White scale bar represent 10 nm long. Color scale ranges from zero (blue) to 0.35 (yellow) normalized LDOS. Black dots indicate the position of Fe impurities. {\bf f} Calculated normalized LDOS curves. The curves are taken far from impurities at the locations marked with circles in {\bf d,e}. $\Delta_0$ is the energy of the superconducting gap.}
		\label{fig:YSRtoAG}
\end{figure*}

\begin{figure*}[ht]
	\centering
		\includegraphics[width=0.7\linewidth]{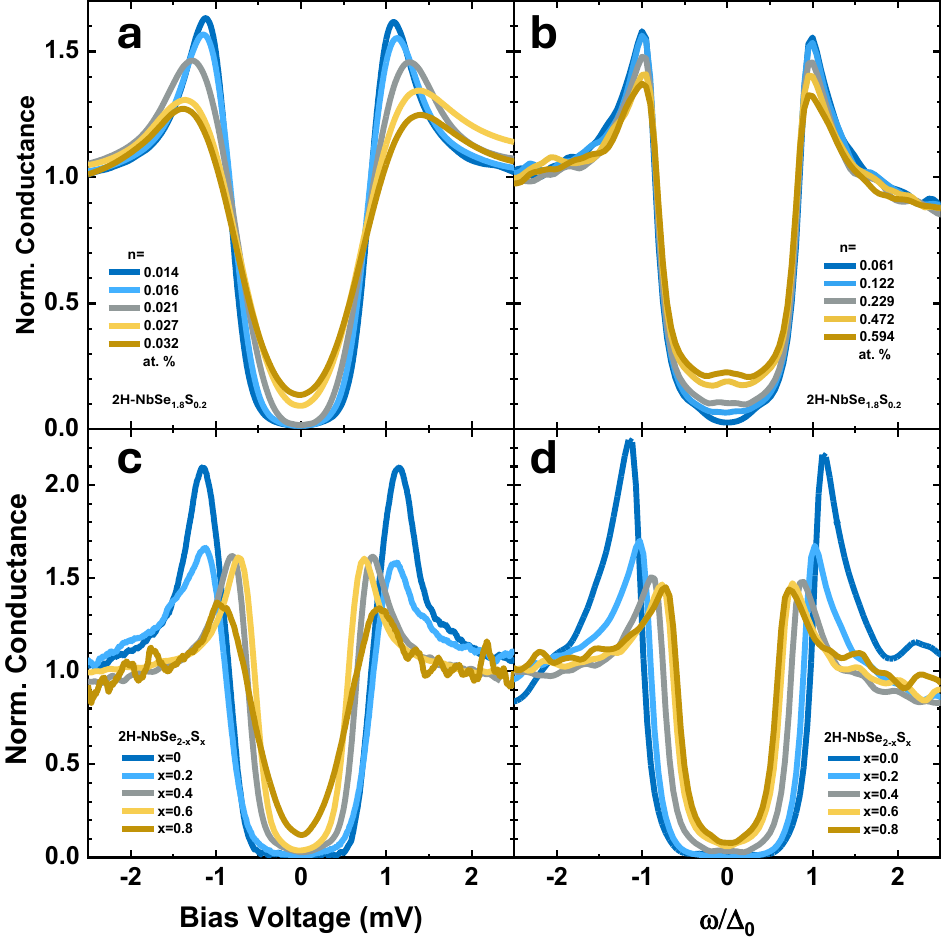} 
	\caption{\ \textbf{Local density of states as a function of magnetic and nonmagnetic impurity concentrations.} {\bf a} Tunneling conductance as a function of the bias voltage at positions far from magnetic impurities and in different fields of view in 2H-NbSe$_{1.8}$S$_{0.2}$ with different magnetic impurity concentrations $n$ ($n=0.014, 0.016, 0.021, 0.027$ and $0.032$ at.\,\% from blue to ocre). The conductance is normalized to its value for bias voltages above the superconducting gap. {\bf b} Calculated normalized LDOS as a function of the energy in 2H-NbSe$_{1.8}$S$_{0.2}$ far from magnetic impurities for different magnetic impurity concentrations ($n=0.061, 0.122, 0.229, 0.472$ and $0.594$ at.\,\%, from blue to ocre). {\bf c} Tunneling conductance as a function of the bias voltage obtained at locations far from magnetic impurities and in fields of view with smallest amount of impurities $n$ in 2H-NbSe$_{2-x}$S$_{x}$ ($x=0,0.2,0.4,0.6,0.8$ from blue to ocre). {\bf d} Calculated normalized LDOS in 2H-NbSe$_{2-x}$S$_{x}$ with nonmagnetic impurity concentrations $x=0.2,0.4,0.6,0.8$ from blue to ocre for a magnetic impurity concentration $n=0.229\%$.}
		\label{fig:gaplessdependence}
\end{figure*}

\begin{figure*}[ht]
	\centering
		\includegraphics[width=0.7\linewidth]{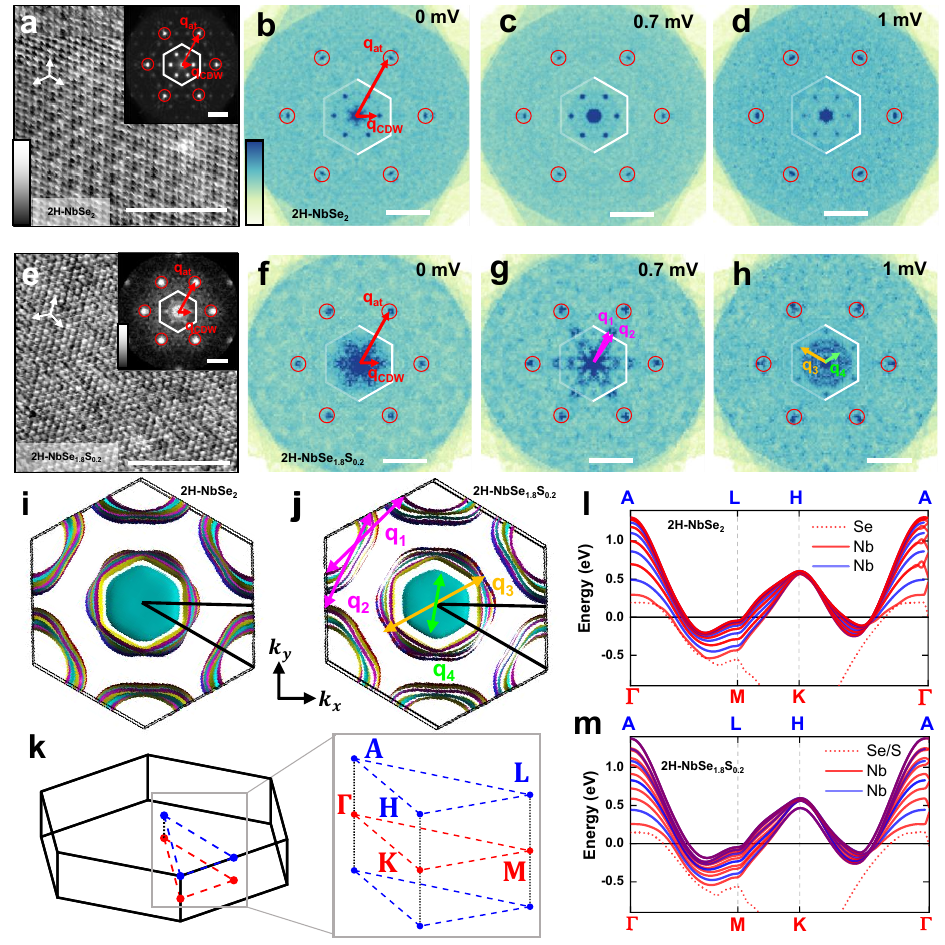} 
	\caption{\ \textbf{States at magnetic impurities and nesting.} {\bf a} ({\bf e}) STM topography image obtained in 2H-NbSe$_2$ (2H-NbSe$_{1.8}$S$_{0.2}$). Insets show the respective Fourier transforms. White scale bars represent 5 nm in real space and 2 nm$^{-1}$ in reciprocal space. White arrows indicate the direction of the three in-plane crystal axes. White to black color scale corresponds to a height difference of $80$ pm. {\bf b-d} and {\bf f-h} Fourier transform of tunneling conductance maps for three different bias voltages ({\bf b,f} 0 mV, {\bf c,g} 0.7 mV and {\bf d,h} 1 mV) obtained in the same fields of view as {\bf a,e}. The tunneling conductance maps for negative bias voltages are shown in Fig.\,\ref{fig:qpiall} and provide the same results. The CDW and atomic lattice Bragg peaks are shown by red arrows. Bragg peaks are also marked with red circles. The white hexagons provide the boundaries of the first Brillouin zone. {\bf i} ({\bf j}) Fermi surface obtained from slab calculations of the band structures of 2H-NbSe$_2$ (2H-NbSe$_{1.8}$S$_{0.2}$). The Brillouin zone and the two main symmetry directions are marked by black lines. Colored arrows in {\bf j} provide the different wave vectors identified in tunneling conductance images in {\bf g,h}. {\bf k} Schematic representation of the first Brillouin zone of 2H-NbSe$_{2-x}$S$_{x}$. We show the main high symmetry points in the grey rectangle. {\bf l} ({\bf m}) Energy versus wavevector dispersion relation in 2H-NbSe$_2$ (2H-NbSe$_{1.8}$S$_{0.2}$). Red (blue) lines are bands derived from Nb orbitals along the $\Gamma$ - M - K - $\Gamma$ line (A - L - H - A). We allow for mixed colors when two curves overlap. This occurs in the upper curves in {\bf m}, which are violet (red+blue) because the bands overlap, evidencing the absence of $k_z$ warping in the Fermi surface sheets for 2H-NbSe$_{1.8}$S$_{0.2}$. Dotted lines provide the Se derived band.}
		\label{fig:qpifft}
\end{figure*}

\clearpage

\section*{Methods}

\subsection*{A. Sample synthesis and characterization}

Large 2H-NbSe$_{2-x}$S$_x$ single crystals ($x=$0, 0.2, 0.4, 0.6, 0.8) were grown by a standard solid state reaction and a subsequent vapor transport process. Crystals were doped with an Fe impurity concentration of $\sim$150 ppm for all samples (corresponding to $n\approx$ 0.020 at.\,\% accounting with experimental uncertainty). The Fe concentration was measured after the growth by inductively coupled plasma analysis. The concentration of other magnetic dopants was found to be $<$25 ppm using inductively coupled plasma spectroscopy.

In Table \ref{tab:caracterizacion} we show the results of the macroscopic characterization of the samples. Elemental analysis from energy-dispersive X-ray spectroscopy (EDS) shows good agreement with the expected stoichiometry from the growth. By refining the powder x ray diffraction (XRD), we obtain the lattice parameters and find that there is a small reduction in both $c$ and $a$ lattice dimensions with increasing $x$. Diffraction peaks agree with the hexagonal 2H structure (space group P63/mmc). Superconducting parameters in Table \ref{tab:caracterizacion} are obtained from resistivity.

\vspace{10pt}

\renewcommand{\tablename}{Extended Data Table}

\begin{table*}[h]
\caption{Lattice constants and superconducting $T_c$ of single crystalline 2H-NbSe$_{2-x}$S$_x$.}
\label{tab:caracterizacion}
\begin{tabular}{cllllll}
\hline
\multicolumn{2}{c}{x   in NbSe$_{2-x}$S$_x$}                   & 0                           & 0.2                         & 0.4                         & 0.6                         & 0.8                         \\
\hline
\multicolumn{2}{c}{a=b (nm)}                           & 0.34451(3)                  & 0.34327(5)                  & 0.34197(3)                  & 0.34099(4)                  & 0.33984(4)                  \\
\multicolumn{2}{c}{c (nm)}                             & 1.2542(1)                   & 1.2506(2)                   & 1.2474(2)                   & 1.2423(2)                   & 1.2387(2)                   \\
\multicolumn{2}{c}{T$_{c}$(K)}                              & 7.2                         & 6.2                         & 5                           & 4.6                         & 3.3                         \\
\end{tabular}
\end{table*}

\subsection*{B. Scanning Tunneling Microscopy}
We used a home-built STM thermally attached to a dilution refrigerator capable of resolving features in the tunneling conductance down to about 8 $\mu$V\,\cite{energyresolution,lab22t}. We used Au tips, cleaned and prepared in-situ as described in Ref.\,\cite{Rodrigo2004} and methods described in Ref.\,\cite{MONTOYA2019e00058}. We use our own measurement and image treatment software, described in Ref.\,\cite{labsoftware}, complemented with other available software\,\cite{Horcas07}. Clean atomically flat surfaces larger than 1 $\mu$m were obtained by cleaving the samples in-situ by gluing an alumina piece to the surface and removing it at 4.2 K using a movable mechanically operated sample holder described in Ref.\,\cite{Suderow2011}. The exposed surfaces are Se/S hexagonal surfaces. Fourier transforms of real space maps have been symmetrized following the C6 symmetry of the surface lattice.
It is important to stress that we make the measurements at temperatures far below the critical temperature $T_c$ (at $<0.1T_c$), which varies in 2H-NbSe$_{2-x}$S$_x$ with $x$ as shown in Table \ref{tab:caracterizacion}.

We have followed the tunneling conductance as a function of temperature in 2H-NbSe$_{2-x}$S$_{x}$ for $x>0.2$. The results are shown in Fig.\,\ref{fig:TC}. We can find the density of states at each temperature by deconvoluting the Fermi function and obtain the result shown in the right panels of Figs.\,\ref{fig:TC}{\bf a,b}. From the density of states we find the superconducting gap as a function of temperature, obtaining a temperature dependence compatible with BCS theory (Fig.\,\ref{fig:TC}{\bf c}).

To estimate the local atomic concentration of magnetic impurities we count the number of in-gap states we observe in a given field of view (marked with dots in Fig. \ref{fig:YSRtoAG} and Fig. \ref{fig:FOV}) and divide by the number of Nb atoms at the surface layer. The number of Nb atoms is calculated by dividing the surface area of a given field of view with the area of a unit cell (there is one Nb atom per unit cell).

\subsection*{C. Numerical modelling of the superconducting LDOS}

To analyze the effect of both magnetic and nonmagnetic impurities we start from the standard BCS real space Hamiltonian given by
\begin{equation}
    H =-\sum_{\iv,\jv,\sigma}t_{\iv\jv}c_{{\iv}\sigma}^{\dagger}c_{{\jv}\sigma} - \mu\sum_{{\iv},\sigma}c_{{\iv}\sigma}^{\dagger}c_{{\iv}\sigma} + \sum_{{\iv}}\Delta_{{\iv}}(c_{{\iv}\uparrow}^{\dagger}c_{{\iv}\downarrow}^{\dagger} + \rm{H.c.}),
\end{equation}
where $c^\dagger_{\iv,\sigma}$ ($c_{\iv\sigma}$) corresponds to the electronic creation (annihilation) operator, with $\iv$ denoting the coordinates of a two-dimensional triangular lattice and $\sigma$ is the spin.
For the kinetic term, we consider up to fifth nearest neighbor hoppings with the parameters $\{ t_1, t_2, t_3, t_4, t_5\} = \{-23, -128.75, -2.2, 7.5, -3 \}$ and chemical potential $\mu = -275$. All parameters considered in the numerical calculations are in units of meV. The crystalline structure contains two NbSe$_2$ layers (Fig.\,\ref{fig:Schematics} {\bf a}), leading to two sets of Nb derived bands (Fig.\,\ref{fig:qpifft} {\bf l,m}). We assume that a single layer is sufficient to capture the main aspects of the in-gap states. This is in line with our DFT calculations showing that the Se-S substitutional disorder makes the bands more two-dimensional.

The magnetic Fe impurities are implemented as
\begin{equation}
    H_{\rm m} = \sum_{\iv} \delta_{\iv \rv_{\rm m}} V_m (c^\dagger_{\iv \up} c_{\iv \up} - c^\dagger_{\iv\down} c_{\iv \down}),
\end{equation}
where $\rv_{\rm m}$ correspond to the positions of the magnetic impurities on the lattice.
The Se-S substitutional disorder is modelled by nonmagnetic impurities included as  repulsive local potentials at random locations, 
\begin{equation}
    H_{\rm p} = V_p \sum_{\iv, \sigma} \delta_{\iv \rv_{\rm p}} c^\dagger_{\iv \sigma} c_{\iv \sigma},
\end{equation}
with $\rv_{\rm p}$ the lattice positions of the nonmagnetic potentials. 

To relate the number of nonmagnetic impurity sites $\sum_i \delta_{i,\bf{r_p}}$ to the experimental S concentration, $x$, we note that the system size $N$ amounts to a number $N$ of unit cells of 2H-NbSe$_2$. Therefore we relate the experimental quantity of Se-S substitution x and the concentration of nonmagnetic disorder in the theoretical model by: \,$x=2\frac{\sum_i \delta_{i,\bf{r_p}}}{N}$.

We construct the $2N^2 \cross 2N^2$ Bogoliubov-de Gennes (BdG) Hamiltonian using the spinor $\Psi_\iv^\dagger = (c_{\iv\uparrow}^\dagger, c_{\iv\downarrow})$, with $N$ denoting the system size in real space.
We introduce the BdG transformation
\begin{equation}
    \begin{cases}
    c^\dagger_{\iv \sigma} = \sum_{n=1}^{N^2} (u_{\iv \sigma}^{n^*} \gamma_n^\dagger + v_{\iv n}^n \gamma_n), \\
    c_{\iv \sigma} = \sum_{n=1}^{N^2} (u_{\iv \sigma}^n \gamma_n + v_{\iv \sigma}^{n^*} \gamma_n^\dagger),
    \end{cases}
    \label{eq:BdG_transf}
\end{equation}
where $(u_{\iv \up}^n, v_{\iv \down}^n)^\top $ denote the set of eigenvectors that diagonalize the BdG Hamiltonian with eigenvalues $E_n$, and hence the diagonalized Hamiltonian is given by $H=\sum_n E_n \gamma_n^\dagger \gamma_n + E_0$. Here, the sum over $n$ includes only the positive eigenenergies. Using the previous transformation, we can solve self-consistently for the superconducting order parameter at each site,
\begin{equation}
    \Delta_\iv = V_{\rm SC} \left( \la c_{\iv\up} c_{\iv\down} \ra - \la c_{\iv\down} c_{\iv\up} \ra \right),
\end{equation}
where $V_{\rm SC}$ is the on-site attractive pairing interaction. In the numerical calculations, we choose $V_{\rm SC} = -90$~meV, which in the homogeneous case gives rise to a gap amplitude $\abs{\Delta_0}\sim 5$~meV. We normalize the energy  $\hbar\omega$  to $\abs{\Delta_0}$ in the figures and we omit $\hbar$ for clarity.

The site-resolved LDOS for a frequency $\omega$ is obtained from the imaginary part of the retarded Green's function,
\begin{equation}
    N_{\sigma}(\iv,\omega) = -\frac{1}{\pi}\Im G_{\sigma}(\iv, \omega),
\end{equation}
where $G_{\sigma}(\iv, \omega)$ is calculated from $G_{\sigma} (\iv, i\omega_n) = - \int_0^\beta d\tau e^{i\omega_n \tau} \la T_{\tau} c_{\iv \sigma} (\tau) c_{\iv \sigma}(0)\ra$ by taking $i\omega_n \rightarrow \omega + i\eta$, with $\eta$ a small parameter that will determine the numerical resolution of the gap. Using the BdG transformation introduced in Eq.~\eqref{eq:BdG_transf}, the total LDOS can be written as
\begin{equation}
    N(\iv,\omega) = -\frac{1}{\pi} \Im \sum_{n,\sigma} \left( \frac{\abs{u_{\iv\sigma}^n}^2}{\omega - E_n + i\eta} + \frac{\abs{v_{\iv\sigma}^n}^2}{\omega + E_n + i\eta} \right).
    \label{eq:LDOS_realspace}
\end{equation}

We normalize the LDOS to its average from $\omega\approx-2\Delta_0$ to $\omega\approx-3\Delta_0$. We introduce a broadening $\eta$ of approximately 2\% $\Delta_0$ to account for the effects due to the finite size of the calculations. This leads, even for pure 2H-NbSe$_2$, to a small finite value of the LDOS inside the gap. The blue curve in Fig.~\ref{fig:S1} {\bf a} displays the spatially averaged LDOS for 2H-NbSe$_{2}$. The ocre curve shows the LDOS at the site of a magnetic impurity. Figs.~\ref{fig:S1} {\bf b-d} show the real-space dependence of the LDOS at $\omega=0$ meV and at a bound state energy $\omega=\pm0.2\Delta_0$. The LDOS around a magnetic impurity has a six-fold shape in real space, which was discussed previously in Refs.~\cite{Menard2015,Uldemolins2022,Uldemolins2024}.

The magnetic potential $V_m$ is fixed by the chosen YSR bound state energies in the single-impurity case, see Fig.~\ref{fig:S1} {\bf a}. We have performed extensive parameter-dependence studies for varying $V_p$, magnetic disorder configurations, and also checked the dependence on $\mu$. Overall, the results presented here are representative of all the performed studies. We note that the dependence of the spatially-averaged LDOS versus $x$ shown in Fig. \ref{fig:gaplessdependence}(d) of the main text is not exhibited at all $\mu$. Only band fillings where $x$ itself causes sufficient changes (lowering) of the DOS near the Fermi level reproduce the experimental results in this respect. In Fig.~\ref{fig:DOSnorm}{\bf a}, we show the non-normalized DOS for the cases of nonmagnetic impurity concentrations $x=0.2,0.4,0.6,0.8$, with a magnetic impurity concentration of $n = 0.229\% $, $V_m = 75$ meV and $V_p = 175$ meV. From these plots, it is evident that the normal state DOS near the Fermi level is reduced upon increasing $x$. The origin of this effect can be traced to a smeared van Hove singularity in the simplified tight-binding band.

In Fig.~\ref{fig:DOSnorm}{\bf b,c,d} we show the effect of increasing $x$ in the zero energy density of states as a function of the position for an area without impurities Fig.~\ref{fig:DOSnorm}{\bf b}, an area with one impurity Fig.~\ref{fig:DOSnorm}{\bf c}, and an area with four impurities, Fig.~\ref{fig:DOSnorm}{\bf d}. To visualize directly the effect of disorder in the zero energy density of states in the superconductor, the DOS is normalized to its value at energies well above the gap in Fig.~\ref{fig:DOSnorm}{\bf b,c,d}. The effect of disorder on the zero energy DOS is pronounced already in absence of magnetic impurities. In presence of magnetic impurities YSR states lose their characteristic six-fold shape and the increased DOS at zero energy is distributed more widely over the studied area.

We can also visualize that increasing the amount of disorder (Fig.~\ref{fig:S2} {\bf a-d}) modifies the spatial extension of YSR states, distributing a finite density of states at the Fermi level over the whole area, and thus increasing the LDOS in between magnetic impurities. We obtain similar results when we make tunneling conductance maps at magnetic impurities (Fig.\,\ref{fig:S2} {\bf e-g}).

\subsection*{D. Density Functional Theory}

Density Functional Theory (DFT) calculations were performed using the Quantum ESPRESSO package\cite{Giannozzi_2009}. The exchange-correlation energy was described using the Perdew-Burke-Ernzerhof (PBE) functional and the generalized gradient approximation (GGA). Van der Waals interactions between layers were included through the Grimme-D3 dispersion correction scheme and pseudopotentials were extracted from the PSL library.  During relaxation, we set a convergence threshold for total energy of 1$\times$ 10$^{-4}$ Ry, and 1$\times$ 10$^{-3}$ Ry/Bohr for the forces. The system studied is a $1\times 1\times 5$ supercell of 2H-NbSe$_2$, composed by 10 Nb and 20 Se atoms. To simulate the 2H-NbSe$_{1.8}$S$_{0.2}$ compound, two of the twenty Se atoms were substituted by S atoms. We employed an energy cutoff of 40 Ry for wavefunctions and 320 Ry for charge density and the Brillouin zone was sampled using a $16\times 16 \times 3$ Monkhorst-Pack k-point mesh, centered at $\Gamma$, which ensures convergence of the total energy and electronic properties.

\setcounter{figure}{0}
\renewcommand{\thefigure}{ED~\arabic{figure}}

\begin{figure*}[t]
    \centering
    \includegraphics[width=0.8\linewidth]{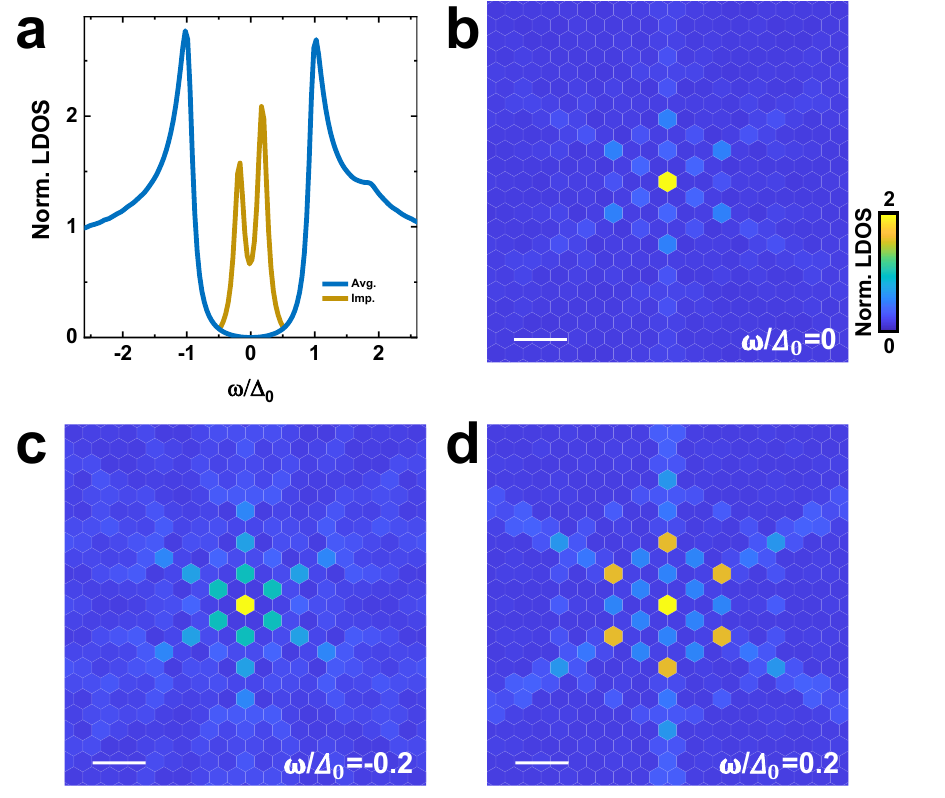}
    \caption{{\bf\ Calculated LDOS around a magnetic impurity.} {\bf a} The calculated LDOS in absence of magnetic impurities is shown as a blue curve and on top of a single magnetic impurity as an ocre curve (we use $V_m = 75$ meV). The systems is $81\cross 81$ sites in size. {\bf b-d} Calculated maps of the LDOS with an impurity located exactly at the center. The LDOS follows the color scale shown in the lateral bar, from large (yellow) to small (blue). White scale bars represent 1 nm {\bf b} is at the Fermi level and {\bf c,d} are obtained at the bound state energies $\omega=\pm0.2\Delta_0$ (peaks in the ocre curve in {\bf a}).}
    \label{fig:S1}
\end{figure*}

\begin{figure*}[ht]
	\centering
		\includegraphics[width=0.99\linewidth]{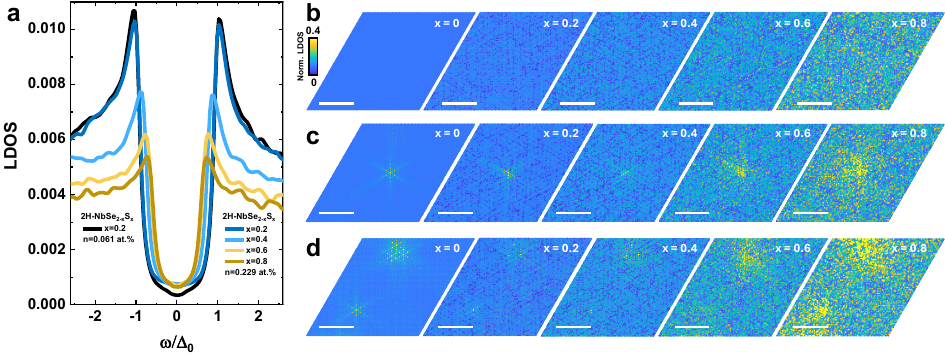} 
	\caption{{\bf\  Non-normalized and normalized local density of states for different nonmagnetic impurity concentrations.} ({\bf a}) Calculated non-normalized LDOS as a function of energy with nonmagnetic impurity concentrations $x=0.2,0.4,0.6,0.8$ from blue to ocre, for a magnetic impurity concentration $n = 0.229\% $. Black line represents $x=0.2$ with $n=0.061 \%$. We consider a system size $81\cross 81$, with $V_m = 75$ meV and $V_p = 175$ meV. A reduction of the normal state DOS and a suppression of the superconducting gap value with $x$ is evident. \textbf{b-d} Site-resolved normalized LDOS with several non-magnetic disorder configurations with no magnetic impurities ({\bf b}), 1 magnetic impurity ({\bf c}) and 4 magnetic impurities ({\bf d}). The color bar is shown at the left. The system size is of $81\cross 81$ unit cells. We use $V_m = 75$ meV, $V_p = 175$ meV, and $\omega=0$ . White scale bars represent 10 nm.}
		\label{fig:DOSnorm}
\end{figure*}

\begin{figure*}[b!]
    \centering
    \includegraphics[width=0.7\linewidth]{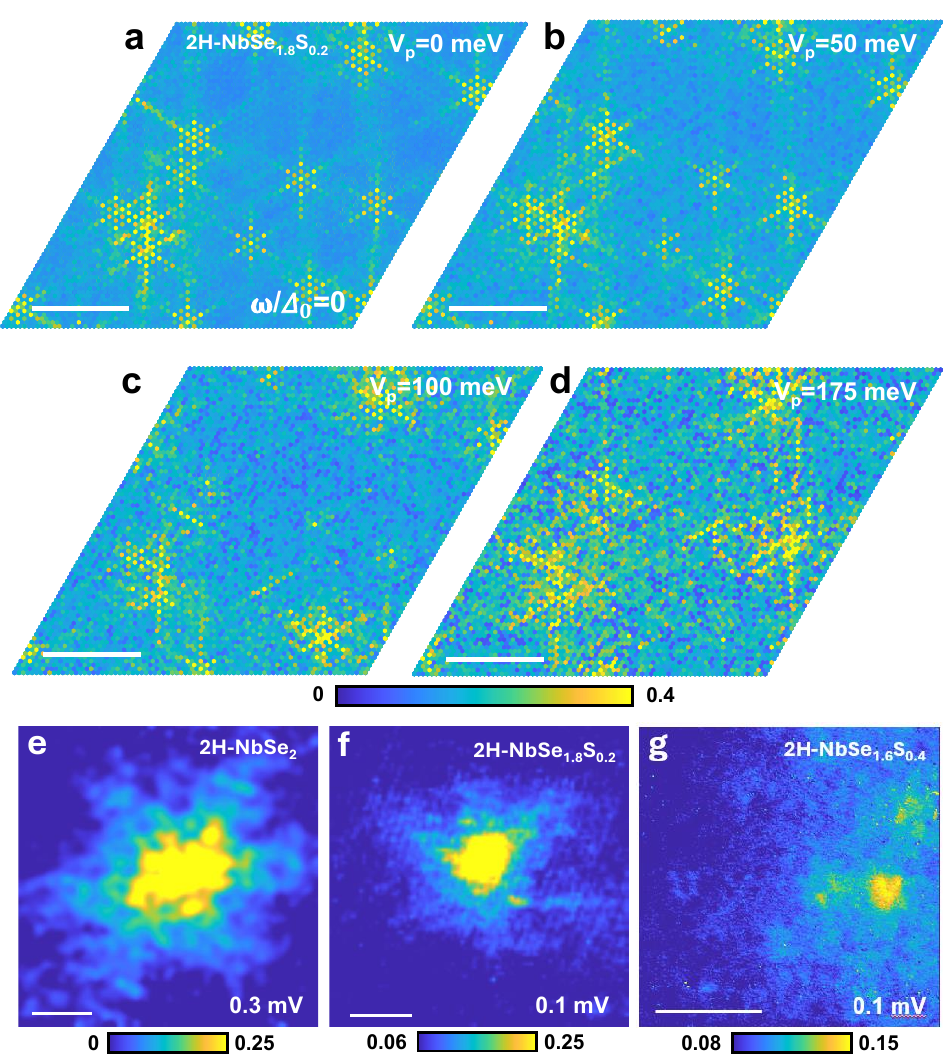}
    \caption{\ \textbf{Influence of non-magnetic disorder on in-gap states}. Site-resolved normalized LDOS with nonmagnetic disorder, considering a magnetic impurity concentration $n = 0.061\%$ and a nonmagnetic impurity concentration $x= 0.2$ in {\bf b-d}. The system size is of $81\cross 81$ unit cells, $V_m = 75$ meV and {\bf a} $V_p = 0$ meV, {\bf b} $V_p = 50$ meV, {\bf c} $V_p = 100$ meV and {\bf d} $V_p = 175$ meV. White scale bar represents 10 nm. {\bf e-g} Tunneling conductance maps at isolated magnetic impurities for {\bf e} 2H-NbSe$_{2}$, {\bf f} 2H-NbSe$_{1.8}$S$_{0.2}$, and {\bf g} 2H-NbSe$_{1.6}$S$_{0.4}$. White scale bar represent 2 nm.}
    \label{fig:S2}
\end{figure*}

\begin{figure*}[ht]
	\centering
		\includegraphics[width=0.9\linewidth]{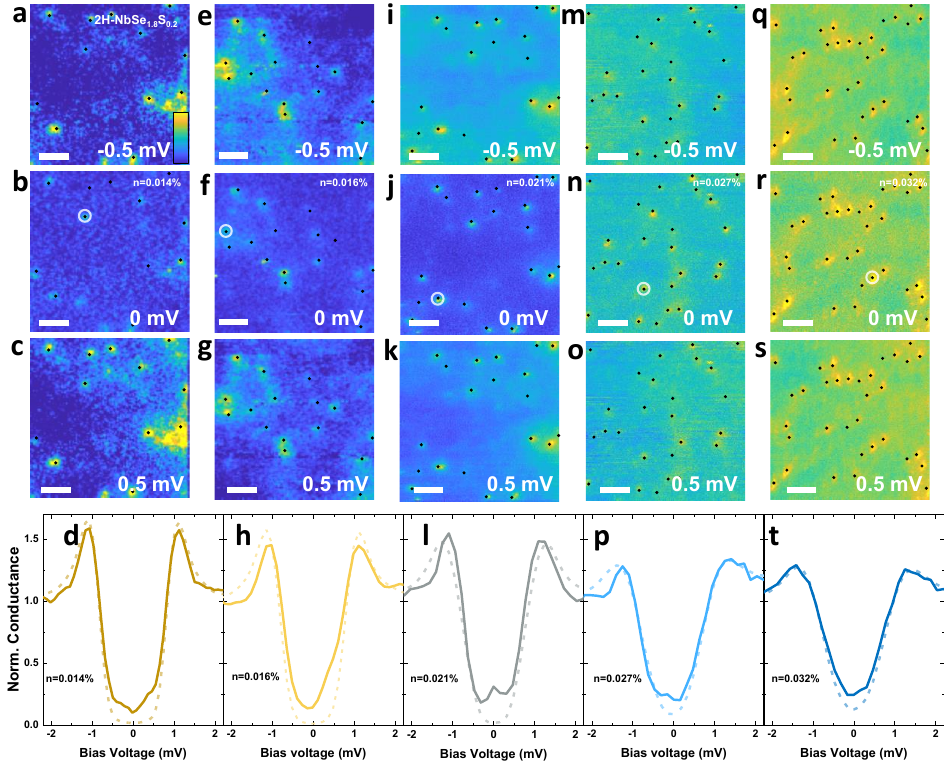} 
	\caption{\ {\bf Tunneling conductance maps at different magnetic impurity concentrations.} Tunneling conductance maps made in 2H-NbSe$_{1.8}$S$_{0.2}$ in fields of view with different $n$, at $-0.5$ mV ({\bf a,e,i,m,q}), $0$ mV ({\bf b,f,j,n,r}) and $+0.5$ mV ({\bf c,g,k,o,s}). Black dots provide the positions of Fe impurities. The tunneling conductance vs bias voltage curves at the impurities marked by a white circle is shown as colored lines in {\bf d,h,l,p,t}. Curves away from impurities are also shown in dashed colored lines. The local magnetic impurity concentration $n$ is 0.014$\%$ in {\bf a-d}, 0.016$\%$ in {\bf e-h}, 0.021$\%$ in {\bf i-l}, 0.027$\%$ in {\bf m-p} and 0.032$\%$ in {\bf q-t}. White scale bars represent 20 nm. The conductance follows the color scale in {\bf a}.}
		\label{fig:FOV}
\end{figure*}

\begin{figure*}[ht]
	\centering
	\includegraphics[width=0.8\linewidth]{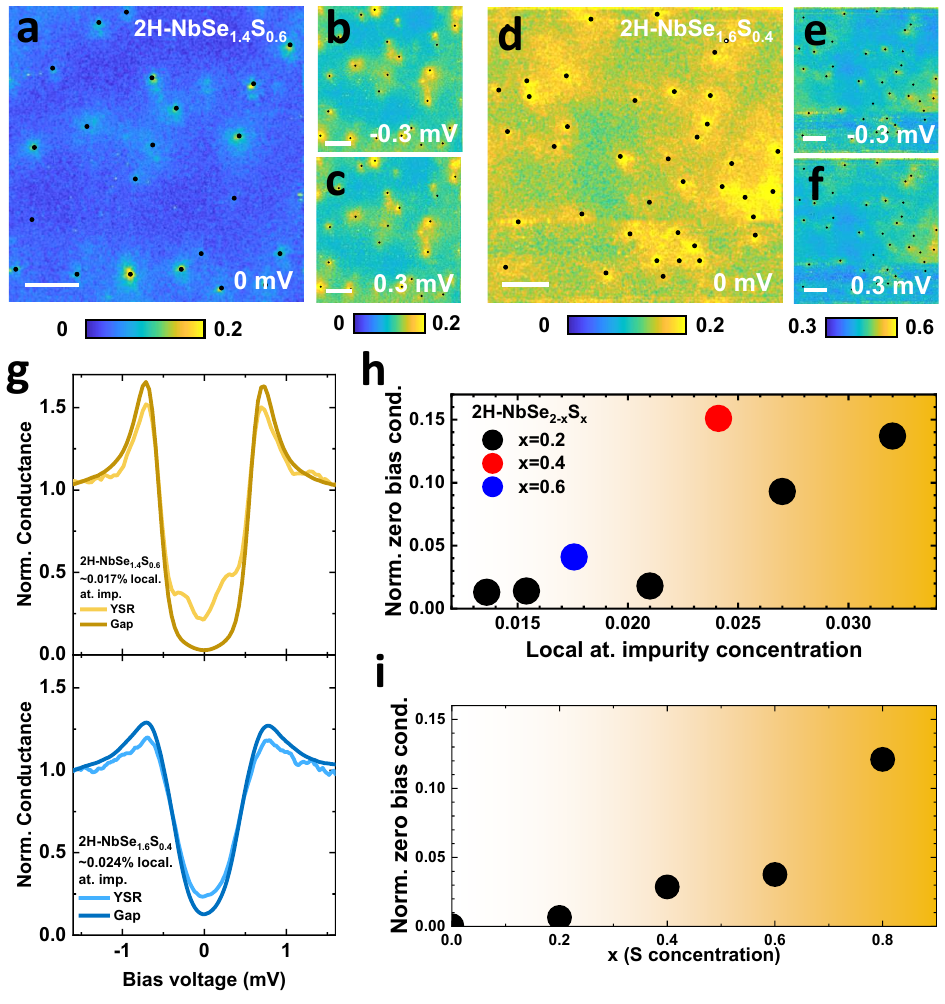} 
	\caption{\ {\bf Results in 2H-NbSe$_{1.4}$S$_{0.6}$ and in  2H-NbSe$_{1.6}$S$_{0.4}$.} {\bf a-c} Tunneling conductance maps in 2H-NbSe$_{1.4}$S$_{0.6}$ with $n=0.017$ at.\% magnetic impurity concentration, at zero bias {\bf a} and at $\pm 0.3$ mV {\bf b,c}. Tunneling conductance maps in 2H-NbSe$_{1.6}$S$_{0.4}$, with $n=0.024$ at.\%, at zero bias {\bf d} and at $\pm 0.3$ mV {\bf e,f}. The conductance follows the color scale given in {\bf a}. White scale bars represent 20 nm. Black dots mark the position of magnetic impurities. {\bf g} Normalized tunneling conductance curves as a function of the bias voltage obtained on top of a magnetic impurity (light colors) and far from magnetic impurities (dark colors) in 2H-NbSe$_{1.4}$S$_{0.6}$ (bottom panel) and in 2H-NbSe$_{1.6}$S$_{0.4}$ (top panel). {\bf h} Normalized zero bias conductance as a function of the concentration of impurities $n$ in 2H-NbSe$_{2-x}$S$_{x}$. {\bf i} Normalized zero-bias conductance for $n \sim0.2$ at.$\%$ as a function of $x$.}
		\label{fig:azufres}
\end{figure*}

\par
\begin{figure*}[ht]
	\centering
		\includegraphics[width=0.6\linewidth]{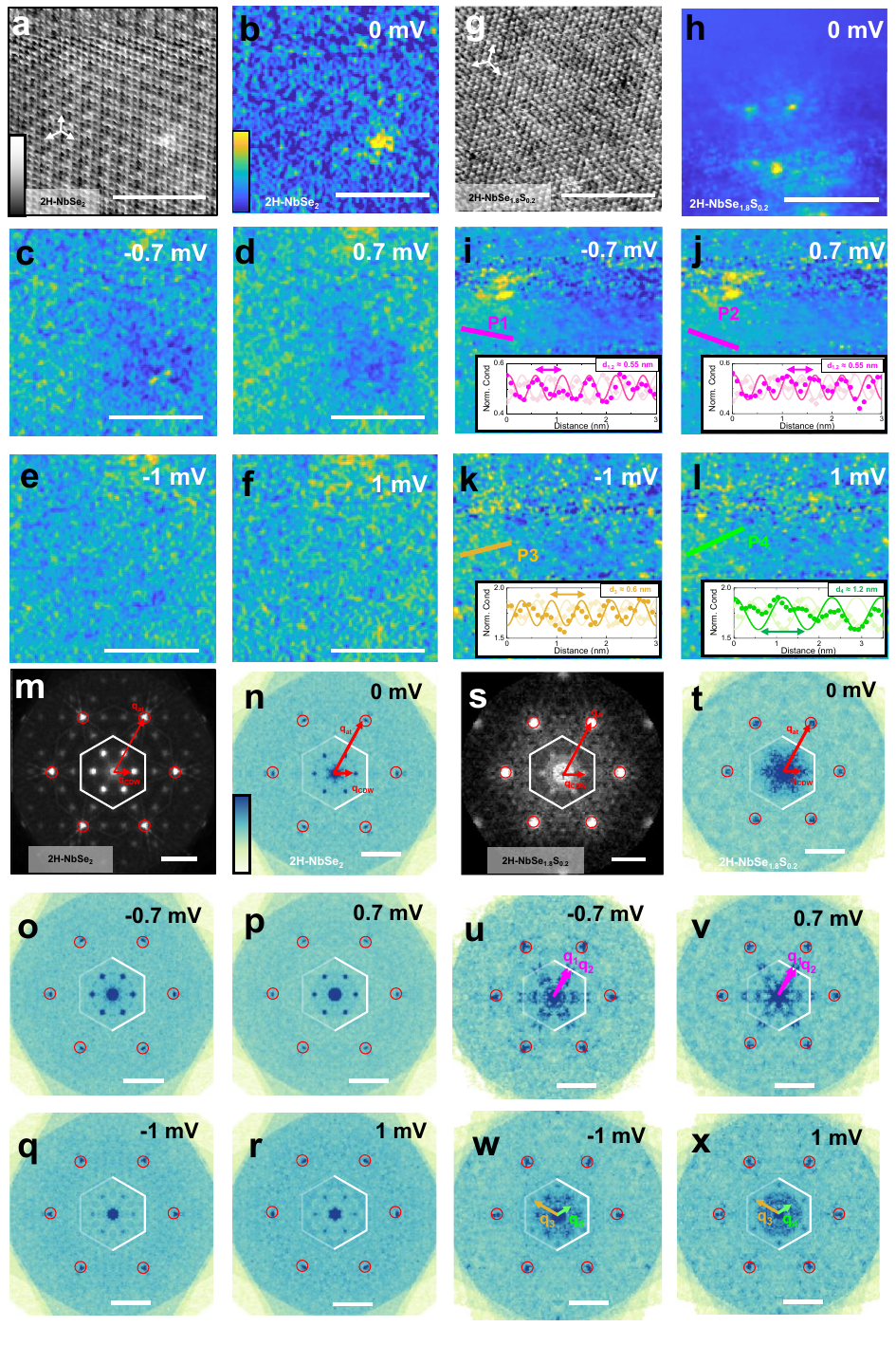} 
	\caption{\ \textbf{In-gap tunneling conductance around magnetic impurities.} In {\bf a,g} we show the STM topography. White scale bars represent 5 nm large in real space and 2 nm$^{-1}$ in reciprocal space. In {\bf b-f} and {\bf h-l} we show the tunneling conductance maps for the bias voltages provided in each panel. Insets in {\bf i-l} provide line scans made along the colored lines as colored points. Light (dark) colored points provide the line scans of the conductance at positive (negative) bias voltage. $P_{1-4}$ stands for for $q_{1-4}$ respectively in Fig.\,\ref{fig:qpifft} {\bf g,h,j}. Lines are oscillations with the wave vectors $q_{1-4}$. In Fig \ref{fig:qpiall} {\bf m,s} we provide the Fourier transform of the topographies in {\bf a,g}, respectively. The CDW and atomic lattice Bragg peaks are shown with red arrows. In {\bf n-r} and {\bf t-x} we show the Fourier transform of the tunneling conductance maps from {\bf b-f,h-l} at the bias voltages shown in each panel. We mark the main wave vectors by colored arrows. The white hexagons provide the boundaries of the first Brillouin zone.}
		\label{fig:qpiall}
\end{figure*}
\par 

\begin{figure*}[ht]
	\centering
	\includegraphics[width=0.8\linewidth]{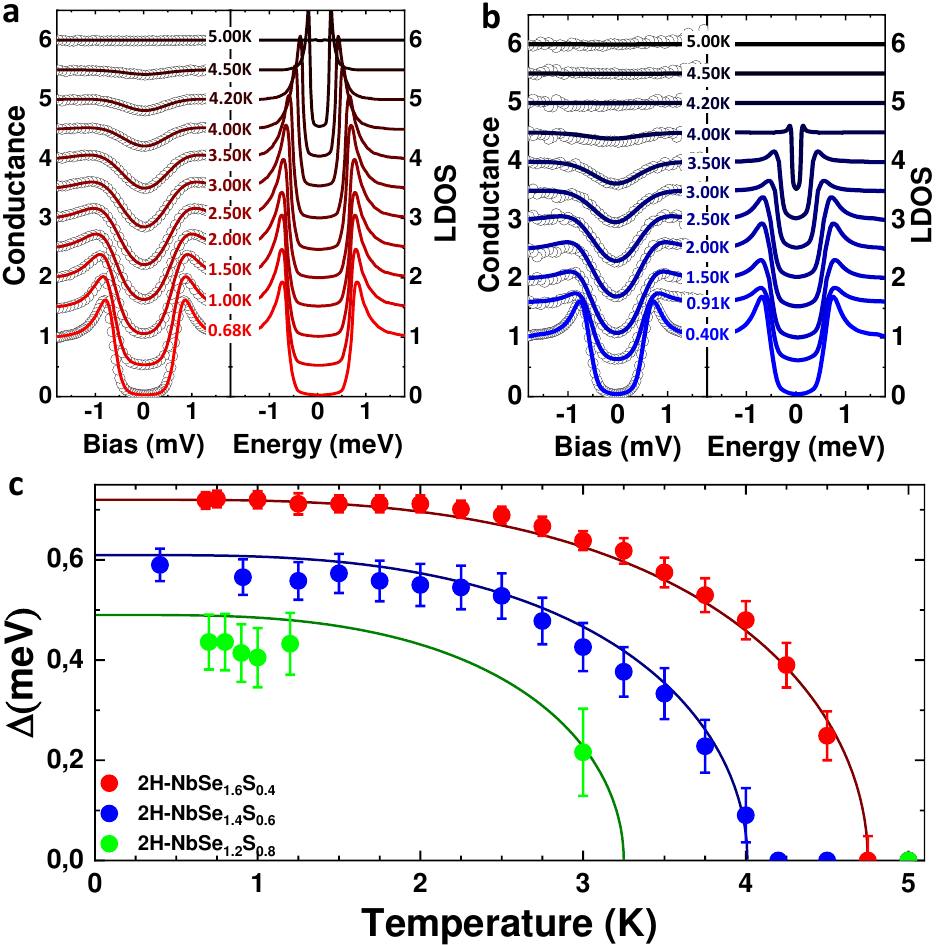} 
	\caption{\ {\bf Superconducting density of states as a function of temperature.} {\bf a,b} Tunneling conductance curves vs bias voltage as a function of temperature in 2H-NbSe$_{1.6}$S$_{0.4}$ and in 2H-NbSe$_{1.4}$S$_{0.6}$. Curves are shifted vertically for clarity. Left panels show the tunneling conductance as points and fits as lines. The fits are made at each temperature using the local density of states shown in the right panels. {\bf c} Temperature dependence of the peaks in the density of states. Red dots are for 2H-NbSe$_{1.6}$S$_{0.4}$, blue dots for 2H-NbSe$_{1.4}$S$_{0.6}$ and green dots for 2H-NbSe$_{1.2}$S$_{0.8}$. Solid lines corresponds to the BCS expression using $\Delta=1.76 \, k_B \,T_c$, with $T_c=4.75 \, K$, $T_c=4 \, K$ and $T_c=3.2 \, K$.}
		\label{fig:TC}
\end{figure*}

\begin{figure*}[ht]
	\centering
		\includegraphics[width=0.8\linewidth]{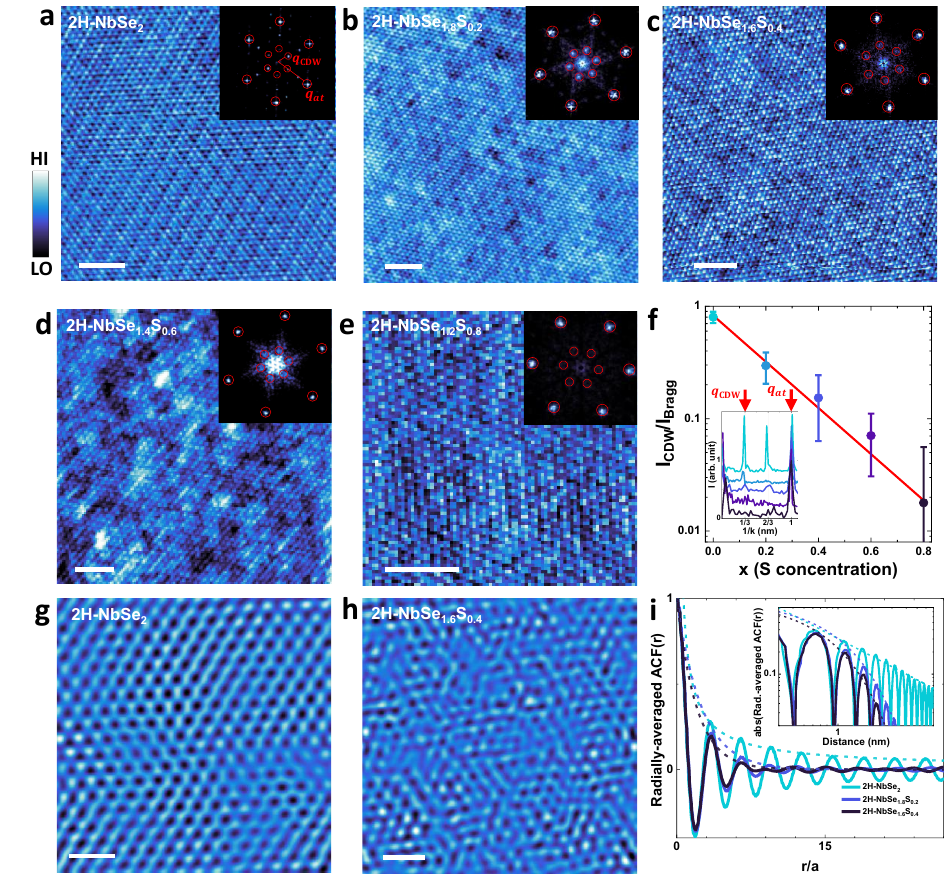} 
	\caption{\ {\bf Atomic scale topography of 2H-NbSe$_{2-x}$S$_x$, $0\leq x \leq 0.8$}. {\bf a-e} STM topography images taken in 2H-NbSe$_{2-x}$S$_x$ for $x=0$({\bf a}), $x=0.2$({\bf b}), $x=0.4$({\bf c}), $x=0.6$({\bf d}), $x=0.8$({\bf e}) (bias voltage of 10 mV and current of 0.8 nA). White scale bars represent 3 nm in size. Insets at the top right show the Fourier transform of the images. Red circles show the CDW and atomic Bragg peaks. In {\bf f} we show the ratio of the intensity in the Fourier transform of the CDW peaks and the atomic Bragg peaks. Red line shows an exponential dependence of the peak height with impurity concentration. In the inset we show the profiles along the CDW and Bragg peaks in the FFT maps from $x=0$ (top curve) to $x=0.8$ (bottom curve). {\bf g,h} CDW real-space oscillations obtained from filtering the CDW modulation in the topography images {\bf a,c}. ({\bf i}) Radially-averaged autocorrelation function $ACF(r)$, with $r$ being the distance normalized to the ab-plane lattice constant $a$ of the CDW (cyan for 2H-NbSe$_{2}$, light blue for 2H-NbSe$_{1.8}$S$_{0.2}$ and black for 2H-NbSe$_{1.6}$S$_{0.4}$). Dashed lines show a exponential decay fit to the $ACF(r)$ for the S-doped materials and a power lay decay for 2H-NbSe$_{2}$. In the inset we show the same data as a function of distance in logarithmic scale.}

		\label{fig:Topos}
\end{figure*}

\begin{figure*}[ht]
	\centering
		\includegraphics[width=0.8\linewidth]{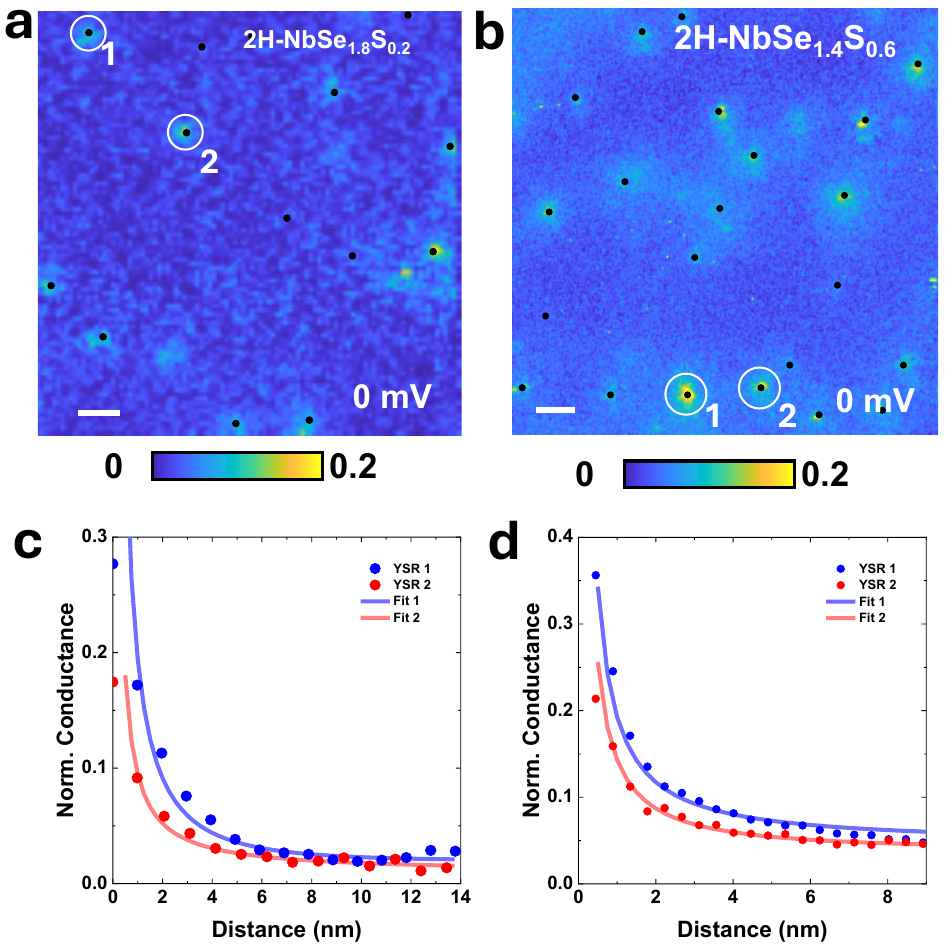} 
	\caption{\ {\bf Spatial dependence of the LDOS around magnetic impurities.} Zero bias conductance maps in 2H-NbSe$_{1.8}$S$_{0.2}$ ({\bf a}) and in 2H-NbSe$_{1.4}$S$_{0.6}$ ({\bf b}). White scale bars represent 10nm. The color scale is shown on the bottom. Magnetic impurities are marked by black dots. {\bf c,d} Radially-averaged tunneling conductance (dots) for the magnetic impurities marked with white circles in {\bf a,b}. Solid lines follow an exponential dependence $\propto e^{-r/r_0}/r \, +C$.}
		\label{fig:asymmetrydecay}
\end{figure*}

\begin{figure*}[ht]
	\centering
	\includegraphics[width=0.7\linewidth]{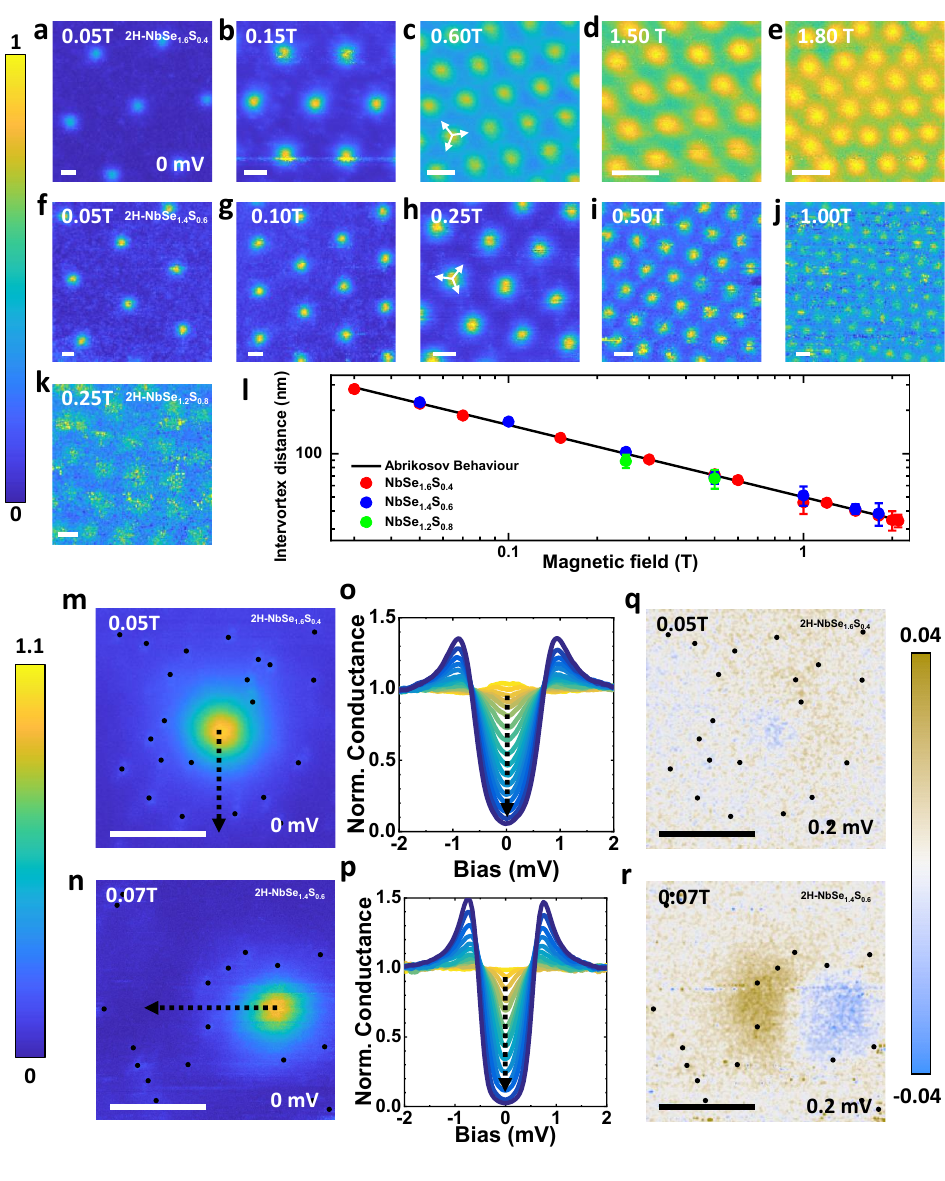} 
	\caption{\ {\bf Vortex lattice and vortex core states in 2H-NbSe$_{2-x}$S$_{x}$.} Tunneling conductance maps at zero bias as a function of the magnetic field in 2H-NbSe$_{1.6}$S$_{0.4}$ ({\bf a-e}), 2H-NbSe$_{1.4}$S$_{0.6}$ ({\bf f-j}), and 2H-NbSe$_{1.2}$S$_{0.8}$ ({\bf k}). Magnetic field values are shown in each panel. The temperature is 400 mK. White arrows indicate the direction of the crystalline directions. The bar at the left provides the color scale in units of the conductance normalized to the value at bias voltages above the gap. Scale bars represent 50 nm. In {\bf l} we show the inter-vortex distance versus magnetic field as red (2H-NbSe$_{1.6}$S$_{0.4}$), blue (2H-NbSe$_{1.4}$S$_{0.6}$), and green (2H-NbSe$_{1.2}$S$_{0.8}$) dots. The black line indicates the Abrikosov prediction for a triangular vortex lattice. In {\bf m,n} we show a zero-bias tunneling conductance map of a single vortex in 2H-NbSe$_{1.6}$S$_{0.4}$ and in 2H-NbSe$_{1.4}$S$_{0.6}$. Magnetic fields are marked in each panel. Fe atoms are marked with black dots. In {\bf o,p} we show the tunneling conductance along the dashed black lines in ({\bf m,n}). In {\bf q,r} we show a map obtained by subtracting tunneling conductance maps at positive from negative bias voltages indicated in each panel ($\pm 0.2$ mV). The color scale is shown at the right.}
		\label{fig:Vortex}
\end{figure*}
\par

\bibliographystyle{naturemag}

\clearpage
\begin{titlepage}

\title{\noindent {\bf \large Supplementary Information of: Gapless superconductivity from extremely dilute magnetic disorder in 2H-NbSe$_{2-x}$S$_x$}}

            \date{}

\baselineskip24pt

\end{titlepage}

\maketitle 

\subsection*{A. Electron-hole asymmetry of the in-gap magnetic states.}

In Fig.\,ED\,4 we show the tunneling conductance maps at different bias voltages in 2H-NbSe$_{1.8}$S$_{0.2}$ in fields of view of approximately 100 nm in lateral size for various local Fe impurity concentrations (14 Fe impurities, $n=0.014$ $\%$, Fig.\,ED\,4{\bf a-d}, 16 Fe impurities, $n=0.016$ $\%$, Fig.\,ED\,4{\bf e-h}, 22 Fe impurities, $n=0.021$ $\%$, Fig.\,ED\,4{\bf i-l}, 28 Fe impurities, $n=0.027$ $\%$, Fig.\,ED\,4{\bf m-p}, 33 Fe impurities, $n=0.032$ $\%$, Fig.\,ED\,4{\bf q-t}). The $+0.5$ mV conductance map with $n=0.014 \%$ (Fig.\,ED\,4\,{\bf c}) shows brighter and more extended YSR states than its $-0.5$ mV counterpart (Fig.\,ED\,4{\bf a}). This situation is reversed for the field of view picked up with $n=0.016 \%$ where we observe a more pronounced hole character in Fig.\,ED\,4{\bf e} compared to Fig.\,ED\,4{\bf g}. As $n$ increases, we observe regions with YSR states with both strong electron and hole features in Fig.\,ED\,4{\bf i,m,q} at $-0.5$ mV and Fig.\,ED\,4{\bf k,o,s} at $+0.5$ mV. The average difference between the density of states at positive and negative bias voltages within the gap and close to magnetic impurities remains of about 15\% for all concentrations of impurities $n$. Thus, the electron-hole asymmetry of YSR states remains unchanged as a function of magnetic impurity concentration and persists well within the range of gapless superconductivity. 

\subsection*{B. Gapless superconductivity and YSR states in 2H-NbSe$_{1.6}$S$_{0.4}$ and in 2H-NbSe$_{1.4}$S$_{0.6}$}

In Fig.\,ED\,5{\bf a-c,d-f} we show tunneling conductance maps in 2H-NbSe$_{1.4}$S$_{0.6}$ and 2H-NbSe$_{1.6}$S$_{0.4}$, respectively, obtained in fields of view of similar size as those discussed in Fig.\,ED\,4. We observe a very similar behavior of the YSR states to those described before in 2H-NbSe$_{1.8}$S$_{0.2}$ (Fig.\,ED\,5{\bf g}), with electron-hole asymmetric in-gap states and the emergence of gapless superconductivity with increasing $x$.

The average tunneling conductance at zero bias  increases strongly with local impurity concentration and is larger for 2H-NbSe$_{2-x}$S$_{x}$ with $x>0.2$ than for $x=0.2$ (Fig.\,ED\,5{\bf h}, where we also show data from Fig.\,3{\bf a}). Indeed, in 2H-NbSe$_{1.6}$S$_{0.4}$ we observe gapless superconductivity already for 0.01$\%$ Fe impurities. 2H-NbSe$_{1.4}$S$_{0.6}$ and 2H-NbSe$_{1.2}$S$_{0.8}$ show gapless superconductivity in all the fields of view that we have analyzed.

\subsection*{C. In-gap states and electronic band structure in 2H-NbSe$_2$ and in 2H-NbSe$_{1.8}$S$_{0.2}$}

In Fig.\,ED\,6({\bf a-l}) we show tunneling conductance maps for several bias voltages in 2H-NbSe$_2$ and 2H-NbSe$_{1.8}$S$_{0.2}$ obtained in the field of view of Fig.\,4({\bf a,e}). The oscillatory behavior is seen on line scans (insets of Fig.\,ED\,6{\bf i-l}). These line scans provide the modulations that lead to the peaks observed in the Fourier transform shown in Fig.\,4. We provide the Fourier transforms for positive and negative bias voltages in Fig.\,ED\,6({\bf m-x}).

\subsection*{D. Charge density wave in 2H-NbSe$_{2-x}$S$_{x}$}

We show the evolution of the CDW peaks in 2H-NbSe$_{2-x}$S$_{x}$ in Fig.\,ED\,8. We observe an exponential suppression of the Bragg peaks of the CDW with increasing $x$ (Fig.\,ED\,8{\bf f}).

Real space maps as those shown in Fig.\,ED\,8 can be written in the form 
\begin{equation}
T(\textbf{r})= \sum_{n=1}^{3} A_{CDW,n}(\textbf{r}) \, e^{i \, \textbf{q}_{CDW,n} \cdot \textbf{r}} +...,
\label{topoec}
\end{equation}
where $\textbf{r}=(x,y)$ is a real space vector, $\textbf{q}_{CDW,n}$ are the three CDW wave vectors, and $A_{CDW,n}$ the amplitude of the oscillations of the CDW\,\cite{lawlerfujita}. All other contributions to the maps are added to the first term of Eq.\,\ref{topoec}. We can now isolate the term $A_{CDW,n}$ by multiplying by $e^{i \, \textbf{q}_{CDW,n}}$, shifting in reciprocal space the origin to the $n^{th}$ peak of the CDW, and multiplying by a Gaussian filter. We get
\begin{equation}
A_{CDW,n}(\textbf{r})= \frac{1}{\sqrt{2 \pi} \sigma} \, \int d\textbf{R} \, T(\textbf{R}) \, e^{i \, \textbf{q}_{CDW,n} \cdot \textbf{R}} \, e^\frac{\left( \textbf{r} - \textbf{R} \right) }{2 \sigma_r^2},
\label{lockinreal}
\end{equation}
where $\sigma_r$ is the cutoff length of the Gaussian filter in real space. $\sigma_r$ is large to include all the information of the CDW signal, but small enough to exclude other patterns. We compute $A_{CDW,n}$ in reciprocal space and then Fourier transform back to real space, 
\begin{equation}
\begin{split}
A_{CDW,n}(\textbf{r})= & F^{-1}[A_{CDW,n}(\textbf{q})]= \\ = &  F^{-1}\left[ F \left(  T(\textbf{r}) \,  e^{i \, \textbf{q}_{CDW,n} \cdot \textbf{r}}    \right) \frac{1}{\sqrt{2 \pi} \sigma}  e^\frac{\textbf{q}^2}{2 \sigma_q^2}  \right],
\label{lockinfft}
\end{split}
\end{equation}
with $\sigma_q=1/\sigma_r$ and  $F[ \, ]$ representing the Fourier transform.

Now, we can trace $T_{CDW,n}(\textbf{r})=A_{CDW,n}(\textbf{r}) \, e^{i \, \textbf{q}_{CDW,n} \cdot \textbf{r}}$ and obtain maps of the spatial-dependent of the amplitude of the CDW, $\left|  A_{CDW,n}(\textbf{r})   \right|$. We then calculate the radially-averaged autocorrelation function $ACF(r)$\,\cite{Postolova2020} and show the result in Fig.\,ED\,8\,{\bf g-i}.

Whereas the $ACF(r)$ in 2H-NbSe$_{1.8}$S$_{0.2}$ and in 2H-NbSe$_{1.6}$S$_{0.4}$ present an exponential decay, in 2H-NbSe$_2$ it shows a power law decay. Thus, we conclude that the long-range coherence CDW order present in 2H-NbSe$_2$ is suppressed by S substitutional disorder, where CDW only occurs within very small nanometer-sized patches.

\subsection*{E. Spatial dependence of the in-gap states at magnetic impurities.}

In Fig.\,ED\,9{\bf a,b} we show maps of the zero bias density of states in 2H-NbSe$_{1.8}$S$_{0.2}$ and in 2H-NbSe$_{1.6}$S$_{0.6}$. In Fig.\,ED\,9{\bf c,d} we show the radially-averaged LDOS around magnetic impurities. We find an exponentially decaying LDOS $\propto e^{-r/r_0}/r \, +C $. The term $C$ accounts for the gapless background. We find $r_0$ of order of a few nm in all cases, showing that magnetic in-gap states are essentially independent from each other.

\subsection*{F. Vortex lattice.}

In Fig.\,ED\,10{\bf a-k} we show the superconducting vortex lattice for 2H-NbSe$_{2-x}$S$_{x}$, $x\leq0.8$ for different magnetic fields applied parallel to the c-axis. We observe in all cases a hexagonal vortex lattice, with the inter-vortex distance closely following the magnetic field dependence from an Abrikosov triangular lattice (Fig.\,ED\,10{\bf l}). We thus see that the substitutional disorder does not influence the vortex lattice symmetry.

We analyze the vortex cores in Fig.\,ED\,10({\bf m-r}). The vortex core bound states characteristic of 2H-NbSe$_2$ are notably suppressed in 2H-NbSe$_{2-x}$S$_{x}$ for $x>0.2$. A small peak in the tunneling conductance at the center of the vortex core remains for x$\le$0.6.

In Ref.\,\cite{victorvortex}, it was shown that the electron-hole asymmetry in the density of states characteristic of the YSR states is transferred to the vortex core bound states. As a consequence, the tunneling conductance maps of vortex cores are electron-hole asymmetric. This provides spatially asymmetric maps when subtracting the tunneling conductance $\sigma$ at positive from the one at negative bias, $\sigma(+V)-\sigma(-V)$. As we show in Fig.\,ED\,10{\bf q,r} the electron-hole asymmetry of the vortex cores remains in 2H-NbSe$_{2-x}$S$_{x}$ for $x>0.2$, in agreement with the magnetic states maintaining their electron-hole asymmetry for $x>0.2$ (Fig.\,ED5{\bf a-f}).

\bibliographystyle{naturemag}

\end{document}